\documentclass[twocolumn,pra,aps,showpacs,showkeys,superscriptaddress]{revtex4}
\usepackage[utf8]{inputenc}
\usepackage{amssymb, amsmath}
\usepackage[dvips,final]{graphicx}
\usepackage[english]{babel}
\usepackage{bm}
\usepackage{subfigure}
\usepackage{color}
\usepackage{ulem}

\newcommand{\ndg}{{\phantom{\dagger}}}
\newcommand{\dg}{\dagger}

\newcommand{\figuresize}{8cm}

\newcommand{\ket}[1]{\left| #1 \right\rangle}

\begin{document}
\author{Julia Kabuss}
\affiliation{Nichtlineare Optik und Quantenelektronik, Institut f\"ur 
Theoretische Physik, Technische Universit\"at Berlin, Hardenbergstr. 36, 10623
Berlin, Germany}
\author{Dmitry O. Krimer}
\affiliation{Institute for Theoretical Physics, Vienna University of Technology,
Wiedner-Hauptstraße 8-10/136, A-1040 Vienna, Austria}
\author{Stefan Rotter}
\affiliation{Institute for Theoretical Physics, Vienna University of Technology,
Wiedner-Hauptstraße 8-10/136, A-1040 Vienna, Austria}
\author{Kai Stannigel}
\affiliation{Institute for Quantum Optics and Quantum Information, Technikerstr.
21, 6020 Innsbruck, Austria}
\author{Andreas Knorr}
\affiliation{Nichtlineare Optik und Quantenelektronik, Institut f\"ur 
Theoretische Physik, Technische Universit\"at Berlin, Hardenbergstr. 36, 10623
Berlin, Germany}
\title{Analytical Study of Quantum Feedback Enhanced Rabi Oscillations}
\author{Alexander Carmele}
\affiliation{Nichtlineare Optik und Quantenelektronik, Institut f\"ur 
Theoretische Physik, Technische Universit\"at Berlin, Hardenbergstr. 36, 10623
Berlin, Germany}
\affiliation{Institute for Quantum Optics and Quantum Information, Technikerstr.
21, 6020 Innsbruck, Austria}
\begin{abstract}
We present an analytical solution of the single photon quantum feedback in a
cavity quantum electrodynamics system based on a half cavity set-up coupled to a
structured continuum. The exact analytical expression we obtain allows us to
discuss in detail under which conditions a single emitter-cavity system, which
is initially in the weak coupling regime, can be driven into the strong coupling
regime via the proposed quantum feedback mechanism [Carmele et al,
Phys.Rev.Lett. 110, 013601]. Our results reveal that the feedback induced
oscillations rely on a well-defined relationship between the delay time and the
atom-light coupling strength of the emitter. At these specific values the
leakage into the continuum is prevented by a destructive interference effect,
which pushes the emitter to the strong coupling limit. 
\end{abstract}
\maketitle
\date{\today}
\selectlanguage{english}

\section{Introduction} 
The basic phenomenon at the heart of any quantum information processing network
is the coherent exchange of photonic and atomic excitations by means of a single
emitter in a microcavity. Advances in the design and fabrication of
microcavities allow very high quality factors and have enabled multiple studies
of cavity quantum electrodynamics (cQED) in the strong coupling limit
\cite{Hennessy:Nature:07a,McKeever:Nature:03,
putz2014_nature_physics,krimer2014_pra,fink_nature_nonlinearity_jcm,
schuster_nature_nonlinearities}. Several applications are proposed and already
realized, such as single-photon transistors, two-photon gateways, parametric
downconversion as well as the generation and detection of individual microwave
photons \cite{chang_nature_transistor,Nielsen::00,monroe_nature_atoms_photons,
zoller-roadmap}. Furthermore, several quantum gate proposals rely on a natural
quantum interface between flying qubits (photons) and stationary qubits (e.g.
atoms). Here, the photons allow for secure quantum communication over long
distances, whereas atoms 
can be used for the manipulation and storage of quantum information
\cite{zoller-roadmap}.
\newline
Applying cQED techniques require that a single-atom/single-photon coupling
exceeds any photon loss and radiative decay processes, such as spontaneous
emission or photon leakage. So, besides technological progress to increase the
quality factor of the cavities, a promising alternative is to identify
strategies to control and exploit potentially advantageous properties of the
environment coupling, which go beyond the conventional effects of the
environment like dissipation and undesired information loss.
%
\newline
A possible mechanism to stabilize qubits and desired quantum states is quantum
feedback based on the repeated action of a sensor-controller-actuator loop. In
such a case, a quantum system is driven to a target state via the external
control \cite{haroche-feedback-control,Wiseman::09}, such that continuous
measurements allow to stabilize the target state, e.g. by a modification of the
pumping strength. In addition to these extrinsic control set-ups, experiments
start to explore a variety of intrinsic, delayed feedback control schemes, e.g.,
by using an external mirror in front of a nanocavity \cite{albert-feedback}.
Intrinsic quantum feedback is not based on a continuous measurement process, but
controls the quantum state by shaping the environment appropriately
\cite{schoell,grimsmo2014rapid,strasberg2013thermodynamics,pyragas2013time,
hein2014optical, schulze2014feedback,Carmele:PhysRevLett:13}. 
%
\newline
Here, we discuss the photon leakage mechanism shaped by an external mirror and
show analytically, how the initial weak atom-cavity coupling is driven into the
strong coupling regime, following Ref.~\cite{Carmele:PhysRevLett:13}. Our
proposed control scheme has potential applications for quantum error correction
\cite{shor-quantum-error-correction}, quantum gate purifying
\cite{zoller-purifying}, or quantum feedback \cite{Wiseman::09}.  
%
\begin{figure}[b!]
\centering
\includegraphics[width=6cm]{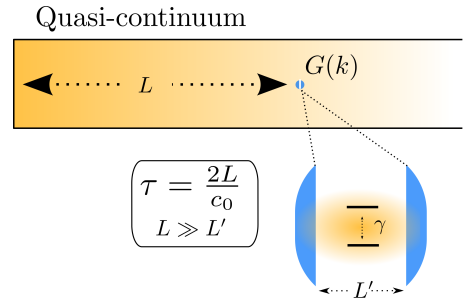}
\caption{Implementation of an intrinsic quantum feedback mechanism via a
quasi-continuum, realized by a photonic crystal waveguide with length $L$,
which is supposed to be considerably larger than the cavity length $L'$. The
waveguide is a half-cavity and allows to exchange cavity photons with waveguide
photons due to the photon leakage $G(k)$. The photons inside the cavity interact
with a single emitter (coupling strength $\gamma$).}
\label{fig:model_schemes}
\end{figure}
\section{Model} 
The system consists of a microcavity system of length $L^\prime$ with a
two-level emitter coupled to a single cavity mode (see Fig.~\ref{fig:model_schemes}). Furthermore, the cavity
exhibits photon loss due to its coupling to external modes. An external mirror,
placed in a distance of $L$, introduces a boundary condition to the external
mode structure and causes a feedback of lost cavity photons into the cavity. We
assume that the microcavity length $L^\prime \ll L$ is very short in comparison
to $L$ to allow a single mode description for the emitter-cavity interaction.
This kind of quantum self-feedback can be realized via a shaped mode continuum
in a photonic waveguide. Due to the finite cavity-mirror distance $L$ and the 
quasi-continuous mode structure of the semi-infinite lead, a
delay mechanism is introduced into the system at $\tau=\frac{2L}{c_0}$ with 
$c_0$ being the speed of light in vacuum. To describe the corresponding physics 
we work with the following 
Hamiltonian within the rotating-wave and dipole
approximation \cite{Walls::07}:
\begin{eqnarray}
\label{eq:Hamiltonian}
H/\hbar 
&=&
-\gamma \left(\sigma^- a^\dg + \sigma^+ a \right)  \\ \notag
& \ & 
- \int \text{d}k \  
G(k,t) \ a^\dg d^\ndg_k + G^*(k,t) \ d^\dg_k a ,
\end{eqnarray}
where a rotating frame is chosen in correspondence to the free energy
contribution of the Hamiltonian. The emitter is described via the Pauli matrices
with $\sigma^{+/-}$ being the raising and lowering operators of the two level
system, respectively. In the following, the atomic energy is assumed to be in
resonance with the single cavity mode.  A photon annihilation (creation) in the
cavity is described with the bosonic operator $a^{\dg}(a)$ and $\gamma$ is the
coupling between the two-level system and the cavity mode. The coupling strength
between the emitter and the field mode is assumed to be of the order of
$\gamma=50~\mu$eV
\cite{McKeever:Nature:03,laucht-dephasing-cavity,berger2014quantum}.  The
cavity photons interact with the external modes $d^{(\dg)}_k$ in front of the
mirror via the tunnel Hamiltonian coupling elements $G(k,t)$. Due to the
rotating frame and the interference with the back-reflected signal from the
mirror, these coupling elements depend both on time $t$ and on the wavenumber
$k$, resulting in the following 
expression for $G(k,t)=G_0 \sin(kL) \exp[i(\omega_0-\omega_k)t]$ with $G_0$
being the bare tunnel coupling strength. $\omega_0$ and $\omega_k$ stand for the
frequencies of a single cavity mode and half-cavity modes, respectively. As we
will see below, this specific form of $G(k,t)$ will determine the nature of the
feedback on the cavity.  
\subsection{Single photon limit}
If no other loss channels or pump mechanism are introduced, the system dynamics
described by the Hamiltonian  Eq.~\eqref{eq:Hamiltonian} can be solved in the
Schr\"odinger picture, following
Ref.~\cite{dorner-half-cavities,cook1987quantum}. In the single
photon limit, the total wave function reads:
\begin{eqnarray}
\label{eq:initial_state}
|\Psi\rangle
&=&
c_e \ket{e,0, \lbrace 0 \rbrace}
+
c_g \ket{g,1, \lbrace 0 \rbrace} \\ \notag
& \ &
+
\int \text{d}k \  
c_{g,k} \ket{g,0, \lbrace k \rbrace },
\end{eqnarray}
where $\ket{e,0, \lbrace 0 \rbrace}$ denotes the excited state of the two-level
system with the cavity and the waveguide being in the vacuum state, $\ket{g,1,
\lbrace 0 \rbrace}$ stands for a single photon residing in the cavity and the
two-level system as well as the radiation field in the waveguide being in the
ground state. Finally, $\ket{g,0, \lbrace k \rbrace }$ describes the ground
state of the two-level system with exactly one photon in the waveguide of mode
$k$. The variables $c_e,\,c_g,\,c_{g,k}$ denote the corresponding amplitudes of
the above three different states. 
\newline
Applying the Schr\"odinger equation, we arrive at the following set of linear
partial differential equations:
\begin{eqnarray}
\label{eq:excited_state}
\partial_t c_e
&=& i \ \gamma \ c_g \\
\label{eq:photon_assisted_ground_state}
\partial_t c_g
&=& 
i \ \gamma \ c_e 
+
i \int \text{d}k \  
\ G(k,t) \ c_{g,k} \\
\label{eq:external_dynamics}
\partial_t c_{g,k}
&=& i \ G^*(k,t) \ c_g.
\end{eqnarray}
First, we numerically solve this coupled set of differential equations assuming
that initially at $t_0=0$ the TLS is in the excited state, $c_e(t_0)=1$, and
there are neither photons inside the cavity, $c_g(t_0)=0$, nor in the external
region, $c_{g,k}(t_0)=0$. To introduce a delay time corresponding to $\tau =
2L/c_0  = 2\pi/\gamma$, we choose a mirror resonator distance $L  =
\pi c_0/\gamma$.

%
%
\begin{figure}[t!]
\centering
\includegraphics[width=\figuresize]{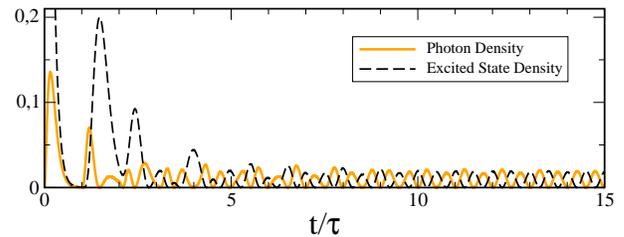}
\caption{Excited state density $|c_e(t)|^2$ of the two-level system (black
dashed line) and the photon density inside the cavity $|c_g(t)|^2$ (orange solid line) in
the quasi-continuum model. The quantum feedback mechanism ($\kappa/\gamma=2$)
induces a regular oscillation pattern at multiples of $\tau=2\pi/\gamma$.}
\label{fig:quasi_continuum}
\end{figure}

The results for the dynamics of the excited state and photon density are shown
in Fig.~\ref{fig:quasi_continuum}. In the time interval  $[0,\tau]$ we find the conventional exponential decay as described by the Wigner-Weisskopf model in
the weak coupling limit. After the first round trip  $\tau=2L/c_0$, the photon
density and after a small delay, also the excited state density, are driven by
the quantum feedback. In this time interval  $[\tau,2\tau]$, the amplitude of
the photon density is smaller than the amplitude of the excited case, since the
damping mechanism acts only on the photons inside the cavity. However, for
longer times, the asymmetry between the amplitudes of the excited state and the
photon density vanishes, so that the system sets into a state of coherent Rabi
oscillations characterized by approximately equal maxima of both densities (see
the asymptotic dynamics for $t/\tau \ge 8$ in Fig.~\ref{fig:quasi_continuum}).
In this long-time limit, the amplitude for the cavity photon population
stabilizes at around 15\% of the maximum photon population
in the first time interval ($[0,\tau]$). This remarkable effect has
been reported in another publication
\cite{Carmele:PhysRevLett:13} and will now be analyzed analytically  and thereby
explained in more detail. In particular, we will focus on the following two
specific questions: (i) How sensitive is this effect on the chosen parameters,
in particular on the choice of the time delay? (ii) Can the oscillation
amplitude be increased or is there an intrinsic limit? To answer these
questions, we will first derive a simplified picture of the dynamics by solving
Eqs.~\eqref{eq:excited_state}-\eqref{eq:external_dynamics} in the Markovian
limit.
%
\subsection{Analytical Quantum Feedback} 
%
The initial decay and the subsequent oscillations observable in Fig.~\ref{fig:quasi_continuum}  
indicates that the underlying physical processes which govern this system consist both of a typical
(Markovian) cavity loss as well as of a (non-Markovian) memory kernel with
significant contributions around multiples of the delay time $\tau$. Assuming that
the rotating wave approximation and the quasi-continuum assumption hold, the
coupling to the external modes can be eliminated from the problem. To achieve
this, Eq.~\eqref{eq:external_dynamics} is integrated formally and inserted into 
Eq.~\eqref{eq:photon_assisted_ground_state} resulting in the following
expression:
\begin{eqnarray}
\label{eq:cg_markov}
\partial_t c_g
&=&
i \ \gamma \ c_e - \kappa \ c_g + \kappa \ c_g \ \Theta(t-\tau) \
e^{i\omega_0\tau},
\end{eqnarray} 
with $\kappa=\pi G_0^2/(2c_0)$.
This reduced expression has the advantage of being easily solvable numerically
and of being amenable to an analytical solution through a Laplace
transformation. With the initial conditions, that neither cavity nor continuum
photons are present in the beginning, i.e.  $c_e(0)=1$, the equations read after
Laplace transformation:
\begin{eqnarray}
\!\!\!\!\!\!\!\!s \ c_e(s) &=& 1 + i \ \gamma \ c_g(s) \label{eq:ces}\\ 
\!\!\!\!\!\!\!\!s \ c_g(s) &=& i \ \gamma \ c_e(s) - \kappa \ c_g(s) + \kappa \
c_g(s) \ e^{-(s-i\omega_0)\tau}, \label{eq:cgs}
\end{eqnarray} 
where $s$ is the complex frequency parameter of the Laplace transformation. As
can be seen from Eq.~(\ref{eq:cg_markov}), the solution consists of a dynamical
component without the mirror induced feedback $t\le \tau$ and one with the
feedback for $t>\tau$.

First, we now derive a solution for the photon-assisted ground state for 
$t\le\tau$, which is the cavity damped Jaynes-Cummings model
\cite{gardiner-book}:
\begin{eqnarray}
c_g(s)=\frac{i \ \gamma}{s^2 + \gamma^2 + \kappa \ s} = \frac{i \ \gamma}{(s
+ \kappa/2)^2 + \gamma^2 - \kappa^2/4 }
\label{eq:cg_laplace} .
\end{eqnarray}
This leads directly to the damped Jaynes-Cummings solutions in the time domain
as expected for times $t\le\tau$, when the cavity-system is not affected by the
feedback mechanism:
\begin{eqnarray}
\label{eq:cavity_damped_jcm}
c_g(t)
&=& 
i \
\frac{\sin\left[ \sqrt{1 - (\kappa/2\gamma)^2} \ \gamma t \right]}
{\sqrt{1 - (\kappa/2\gamma)^2}} \ e^{-\kappa/2 \ t} .
\end{eqnarray} 
Note, due to the cavity damping not only the amplitude is reduced but also the
Rabi oscillation frequency is reduced by a factor of 
$\sqrt{1-(\kappa/2\gamma)^2}$.
The cavity loss leads inevitably to an effectively reduced value for the
coupling strength, and as a result, the frequency of damped Rabi oscillations
decreases. {\it As we will see below, this restriction is lifted if a feedback
mechanism is present}.
\newline
Now, we solve the dynamics for times  $t>\tau$. This leads to an additional term
in the denominator. By using a geometric series expansion, i.e. 
$(1-q)^{-1}=\sum_n q^n$ for  $q<1$ and  $n\rightarrow\infty$,
Eq.~\eqref{eq:cg_laplace} can be written as:
\begin{eqnarray}
c_g(s)
&=&
i \ \gamma 
\frac{\left[1- \frac{\kappa \ s \ \exp[-(s-i\omega_0)\tau]}{(s + \kappa/2)^2 +
\gamma^2 - \kappa^2/4 } \right]^{-1}}{(s + \kappa/2)^2 + \gamma^2 - \kappa^2/4
} 
\\ \notag 
&=&
i \ \gamma
\sum_n
\frac{\left[ \frac{ \kappa \ s \ \exp[-(s-i\omega_0)\tau]}{(s + \kappa/2)^2 +
\gamma^2 - \kappa^2/4 }  \right]^{n}}{(s + \kappa/2)^2 + \gamma^2 - \kappa^2/4
} .
\end{eqnarray}
Due to the linearity of the Laplace transformation, the solution in the time
domain can be obtained via the method of partial fraction expansion and the
convolution property.
%
\begin{figure}[t!]
\centering
\includegraphics[width=\figuresize]{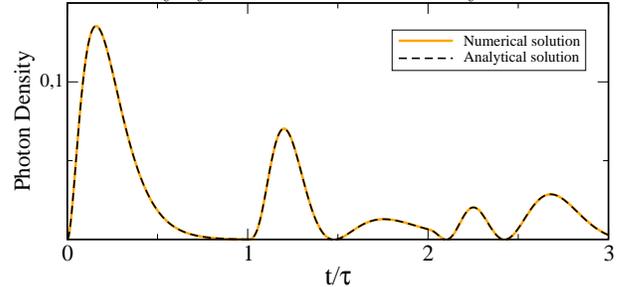}
\caption{Comparison between numerical calculation and the Laplace transformed
analytical solution valid until $t=3\tau$ with $\kappa/\gamma=2$. }
\label{fig:transient}
\end{figure}
%
However, the expression is very lengthy and must be calculated for every time
interval  $[n\tau,(n+1)\tau]$, separately. Since we are interested in the weak
coupling regime, we can choose the parameter to be: $\gamma=\kappa/2$ to
simplify the expression into:
\begin{eqnarray}
c_g(s)\!=\!
\frac{i \ \kappa/2}{(s + \kappa/2)^2} \!
\sum_n \left[\frac{\kappa s e^{(i\omega_0-s)\tau}}{(s + \kappa/2)^2}\right]^n.
\end{eqnarray}
Using now the binomial series and Laplace transformation:  $n!/(s-a)^{n+1}
\rightarrow t^n\exp[at] $, we get an expression in the time domain:
\begin{widetext}
\begin{eqnarray}
c_g(t)
&=&
\frac{i}{2}
\sum_{n=0}^{\infty} n! 2^{n+1} e^{-\kappa/2(t-n\tau)+i\omega_0n\tau} 
\Theta(t-n\tau) 
\sum_{k=0}^{n} \frac{(-1)^{k}}{k!(n-k)!}
\frac{\left[\kappa/2(t-n\tau)\right]^{n+1+k}}{(n+1+k)!}
\label{eq:cg_analytical}.
\end{eqnarray} 
\end{widetext}
In Fig.~\ref{fig:transient}, the numerical solution of
Eq.~\eqref{eq:excited_state} coupled with Eq.~\eqref{eq:cg_markov} is compared with the analytical solution Eq.~\eqref{eq:cg_analytical} for the time interval  $[0,3\tau]$. The excellent agreement found between these two solutions confirms the validity of our calculations. For longer times, more
terms from the series expansion (\ref{eq:cg_analytical}) should be included, which is a straightforward procedure. As a next step, we derive the long time behaviour using the residuum method.

\subsection{Long time solution} 
%
The long time dynamics of the coupled system is directly related to the
singularities in the contour integral of the Laplace transformed function
\cite{Krimer:PhysRevA:14}. To demonstrate this explicitly, we need to find the
singularities of the photon-assisted ground state amplitude:
\begin{eqnarray}
c_g(s)
&=&
\frac{i\gamma}{s^{2}+\gamma^{2}+\kappa s- \kappa s e^{-(s-i\omega_0)\tau}}.
\end{eqnarray} 
The singularities are found by setting the denominator to zero. We assume a pure
oscillation behavior in the long time limit, i.e., where $s$ is purely
imaginary. We set $s=\pm i\gamma$ and get:
\begin{eqnarray}
-\gamma^{2} + \gamma^{2} \pm i \gamma \kappa  \left(1-e^{\mp i \gamma \tau}
e^{i\omega_0\tau}\right)=0,
\end{eqnarray} 
from which immediately follows
\begin{eqnarray}
\label{eq:pure_oscillation_condition}
e^{i (\omega_0 \mp \gamma) \tau} =1.
\end{eqnarray} 
%
%
\begin{figure}[t!]
\centering
\includegraphics[width=\figuresize]{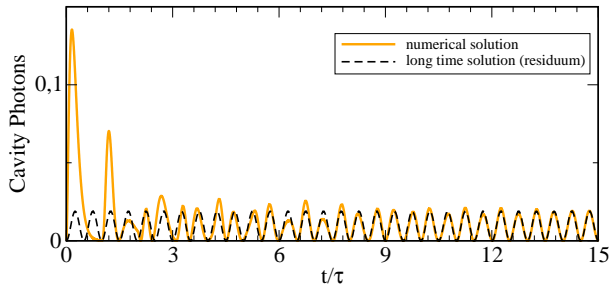}
\caption{Comparison between numerics (orange curve) and long time solution
determined by the residuum contribution only (black curve) for an initially
excited TLS $c_e(0)=1$ and $\kappa/\gamma=2$. 
After several $\tau$, the analytical long time solution and the numerics
coincide. }
\label{fig:long_time_solution}
\end{figure}
%
We now need to find a delay time $\tau$ in a way, that for  $\pm i\gamma$, this
equation is valid. As it turns out, the corresponding singularity condition can
be matched for the following two cases:
\begin{eqnarray}
\label{cond_i}
(i) \exp(i \gamma \tau) = \exp(i \omega_0 \tau)  = 1 \\
(ii) \exp(-i \gamma \tau) =  \exp(i \omega_0 \tau)  = 1.
\label{cond_ii}
\end{eqnarray}
In order to satisfy these conditions we observe that we can freely choose the
delay time with respect to the coupling strength by adjusting the length $L$
between cavity and mirror accordingly. For instance, if we choose $\gamma
\tau=2\pi m$ and at the same time tune the resonance frequency such that
$\omega_0\tau=2\pi l$, where $l,m$ are integer numbers, then the conditions
$(i)$ and $(ii)$ are satisfied. (Note that
there are also three other obvious constraints on $\tau$ and $\gamma$ to satisfy
the conditions $(i)$ or $(ii)$ but they are not discussed below.) As a result we
achieve a purely coherent asymptotic solution with a minimum of dephasing and a
maximum amplitude, corresponding to the fact that the pole does not contain any
decaying term. Indeed, we derive the following expression for the asymptotic
behaviour of the photon density inside the cavity
\begin{eqnarray}
&&c^{(i)}_g(t)
= \frac{1}{2\pi i}
\oint  \text{d}s \ c_g(s) \ e^{st}
\approx\sum_{ \text{Poles}  }  \text{Res}\left[c_g(s) e^{st} \right]\,\,\,\, \\
&=& \sum_\pm \underset{s\rightarrow\pm i\gamma}{\text{lim}} 
\frac{(s\pm i\gamma) \exp[s2 n\pi/\gamma] \
i\gamma}{(s+i\gamma)(s-i\gamma)+\kappa
s\left(1-\exp[-s2 m\pi/\gamma]\right)}.
\nonumber
\end{eqnarray}
Using now L'H\^opital's rule and taking the limits $s\rightarrow\pm i\gamma$,
the solution for $c_g(t)$ reads
\begin{eqnarray}
\label{eq:long_time_solution}
c_g(t)=\frac{i \sin[\gamma t]}{1+\kappa m \pi/\gamma}.
\end{eqnarray} 
In Fig.~\ref{fig:long_time_solution}, the numerical solution and the analytical
long time solution is plotted for $\tau=2\pi/\gamma$ (i.e. when $m=1$) up to
several 
$\tau$. The agreement is excellent with the long time solution accurately
recovering the amplitude and the oscillation frequency of the numerical
solution. Interestingly, for  $\kappa=0$, we recover the Jaynes-Cummings
solution as in Eq.~(\ref{eq:cavity_damped_jcm}) with $\kappa=0$. In contrast to
Eq.~(\ref{eq:cavity_damped_jcm}), we see, however, that the cavity loss does not
modify the frequency of vacuum Rabi oscillations, which is now equal to the coupling
strength $\gamma$, and only damps the corresponding amplitude. In this context,
we discover a maximum amplitude for the feedback effect via this proposed
mechanism. It is seen that Eq.\eqref{eq:long_time_solution} yields the
following amplitude of the quantum feedback induced Rabi oscillations for 
$\tau=2\pi/\gamma$: $1/(1+\kappa\pi/\gamma)^{2}$. Therefore with $\kappa
= 2 \gamma$, the maximum amplitude is approximately  
$0.02$ in correspondence with Fig.~\ref{fig:long_time_solution}, which is 15\%
of the maximum photon amplitude in the first time interval
$[0,\tau]$. With these results at hand we can now also
answer the questions raised above:
\newline
(I) The effect of stabilized Rabi oscillations in the long time limit depends
strongly on the chosen time delay $\tau$, which has to be chosen such to satisfy one of the conditions (\ref{cond_i}), (\ref{cond_ii}) that lead to asymptotically undamped Rabi
oscillations. Furthermore, the factor $\exp[i\omega_0\tau]$
plays a crucial role to decide whether quantum feedback leads to a stabilized
Rabi oscillation or to a damped feedback situation. However, the effect depends
only quantitatively (rather than qualitatively) on the cavity loss $\kappa$ and
coupling strength  $\gamma$, besides the obvious restriction that both of them
are unequal to zero.
\newline
(II) For a given ratio between the coupling strength and the cavity loss,
$x=\kappa/\gamma$, there is a maximum amplitude which is given for the above
case by $(1+x\pi)^{-2}$. 

\subsection{Photon-Path Representation} 

To give an intuitive explanation for this effect of recovered Rabi oscillations
in the weak coupling limit, we visualize 
the resulting cavity dynamics in the framework of the photon-path representation
\cite{Alber:PhysRevA:13,Stampfter:PhysRevE:05}. For this purpose we rewrite the
system of equations of motion \eqref{eq:ces}, \eqref{eq:cgs} in the Laplace domain as: 
\begin{eqnarray} 
\begin{pmatrix}
c_e(0) \\ c_g(0)
\end{pmatrix}
&=&
s
\left[
1
-\mathbb{L}
\right] 
\begin{pmatrix}
c_e(s) \\ c_g(s) 
\end{pmatrix},
\end{eqnarray} 
with the scattering matrix:
\begin{eqnarray}
\mathbb{L} =
\begin{pmatrix}
0 & i\frac{\gamma}{s} \\
i\frac{\gamma}{s} & -\frac{\kappa}{s}\left(e^{-\tau s} -1 \right) \\ 
\end{pmatrix}.
\end{eqnarray}
Assuming that there is an inverse to that matrix
and using the Neumann expansion, we get for $|| \mathbb{L} || < 1$:
\begin{eqnarray}
\begin{pmatrix}
c_e(s) \\ c_g(s)
\end{pmatrix}
&=&
\frac{1}{s}
\sum_{n=0}^\infty
\mathbb{L}^n 
\begin{pmatrix}
c_e(0) \\ c_g(0)
\end{pmatrix} \\
&=&
\sum_{n=0}^\infty
\left[
\frac{(i\gamma)^n}{s^{n+1}}
\begin{pmatrix}
0 & 1 \\ 1 &  \frac{\kappa}{i\gamma} (e^{-\tau s} - 1)
\end{pmatrix}^n
\right]
\begin{pmatrix}
c_e(0) \\ c_g(0)
\end{pmatrix}\nonumber
\end{eqnarray} 
Now, the dynamics is written in a very complicated manner but in a way that the
photon path (represented by scattering processes due to $\mathbb{L}$) becomes
visible. With this expansion, one can represent the dynamics as a series of
single scattering events by multiple application of the matrix, which swap the
excitation from $c_e$ to $c_g$ and includes the cavity loss and the gain from
the feedback. This becomes especially apparent, when writing down the single
terms of the Laplace transform and then transforming them back into the time
domain. In particular for the ground state probability such an expansion reads
(terms up to $t^3$ are kept only):
\begin{eqnarray}
\!\!\!\!\!\!\!\!\!&&c_g(t)=\dfrac{(i\gamma)}{1!}-\dfrac{(i\gamma)\kappa}{2!}
t^2+\dfrac{ (i\gamma)^3}{3!}t^3+\dfrac{(i\gamma)\kappa^2}{3!}t^3+...
\nonumber
\\
\!\!\!\!\!\!\!\!\!&&\dfrac{(i\gamma)\kappa}{2!}
(t-\tau)^2\theta(t-\tau)\!-\!\dfrac{
(2i\gamma)\kappa^2}{3!}(t-\tau)^3\theta(t-\tau)\!+\!\!...\,\,\,\,\,\,\,\,\,\,
\end{eqnarray}
From the structure of this expansion follows that undamped Rabi oscillations can
be viewed as a result of an interference between incoming and outgoing photonic
paths provided that $\tau=2\pi/\gamma$. In other words, the strong coupling
feature is produced by a destructive interference effect of the photon paths at the point
within the waveguide, where the tunneling event between cavity and
waveguide takes place. This expansion explains furthermore that it takes a
minimum time for this effect to unfold, since at least two dissipatively
interacting waves need to be in the waveguide.
%
\subsection{Strong coupling limit} 
%
To complete the picture, we investigate the proposed feedback mechanism via a
quasi-continuum in the strong coupling limit.
\begin{figure}[t!]
\centering
\includegraphics[width=\figuresize]{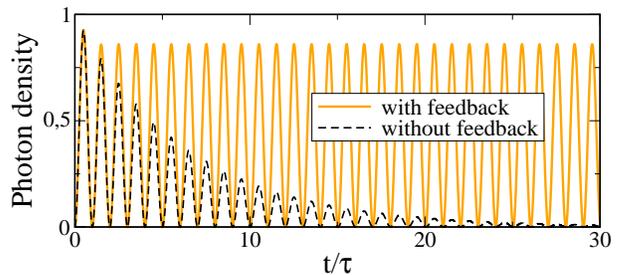}
\caption{Fast stabilization of Rabi oscillations of the cavity photon number in
the strong coupling limit $\gamma = 20 \kappa$ after only one roundtrip $\tau$,
when the feedback from the waveguide is present (orange curve). Initially, the excitation is stored in
the TLS ($c_e(0)=1$). }
\label{fig:strong_coupling}
\end{figure}
%
In Fig.~\ref{fig:strong_coupling}, the dynamics with and without feedback is
plotted for a a coupling strength  $\gamma=20\kappa$.  Clearly, Rabi
oscillations are visible with and without feedback. If no feedback is present, however,
the amplitude of the Rabi oscillations are damped fast without changing the
frequency. With a feedback and a chosen delay time of 
$\tau=\pi/(2\gamma)$, on the other hand, the amplitude loss is 
stopped at very early times already: Already
after one roundtrip the amplitude stays constant for all times, if no other
loss mechanism inside or between mirror and cavity is present.
We explain the acceleration of the stabilization feature by the fact, 
that for the strong coupling regime the in- and outgoing photons
already interfere efficiently after one roundtrip.
In contrast, in the weak coupling limit the in- and out tunneling does
not overlap for the first three roundtrips and as a consequence
inteference takes place at longer times only.
If we choose a larger roundtrip time $\tau$, also in the strong
coupling regime, a higher number of roundtrips $(n\tau)$ is necessary to
reach the point of stabilized Rabi oscillations.
%

\section{Conclusion and outlook} 
We have discussed an approach to stabilize single-emitter cQED via a quantum
feedback mechanism induced by an external mirror. Our analytical solution shows
that depending on the chosen parameters an intrinsic limit of the feedback
effects exists. For a system initially in the weak coupling regime (before the
feedback modifies the system dynamics) we demonstrate that the quantum feedback
can at most recover approximately 15\% of the maximum cavity occupancy in
the first time interval.
Our analytic calculations demonstrate furthermore that the quantum feedback
induced Rabi
oscillations are indeed coherent and follow a typical differential delay
equation with an appropriate inhomogeneity to drive the system into the strong
coupling regime. Our results extend the set of exact analytical solutions in the
field of coherent atom cQED and form a starting point to establish a framework
for a theoretical description of coherent quantum feedback.
We would like to thank N. Naumann and S. Hein for helpful discussions.
We acknowledge support from Deutsche Forschungsgemeinschaft through SFB
910 ``Control of self-organizing nonlinear systems''.
D.K. and S.R. are supported by the Austrian Science Fund (FWF) through project No. F49-P10 (SFB NextLite).
AC acknowledges gratefully support from Alexander-von-Humboldt foundation
through the Feodor-Lynen program.

\end{document}